# Créativité assistée par ordinateur : composer la musique d'un film en utilisant uniquement sa courbe de luminosité extraite automatiquement

**Computational Creativity: Compose the Music for a Movie using only its Automatically Extracted Brightness Curve**

Felipe Ariani[1], Marcelo Caetano[1,3], Javier Elipe Gimeno[1], Ivan Magrin-Chagnolleau[1*,2]

[1]Aix-Marseille Univ, CNRS, PRISM, Marseille, France
[2]Chapman University, California, USA
[3]Schulich School of Music, McGill University, Montreal, Canada

[*]Correspondant : ivan.magrin-chagnolleau@cnrs.fr

**Résumé**

Dès sa conception, l'ordinateur a trouvé des applications pour accompagner la créativité des humains. De nos jours, le débat sur les ordinateurs et la créativité implique plusieurs défis, tels que comprendre la créativité humaine, modéliser le processus créatif, et programmer l'ordinateur pour qu'il présente un comportement qui semble être créatif dans une certaine mesure. Dans cet article, nous nous intéressons à la manière dont l'ordinateur peut être utilisé comme un outil favorisant la créativité dans une composition musicale. Nous avons extrait automatiquement la courbe de luminosité d'un film muet et l'avons ensuite utilisée pour composer une pièce musicale pour accompagner le film. Nous avons extrait plusieurs paramètres de la courbe de luminosité, et nous avons appliqué des règles de composition à partir de ces paramètres pour écrire la musique instrumentale du film. La composition finale présente une synchronicité et un ajustement esthétique avec le film, qui sont surprenants. Ce processus de composition a aussi permis un degré de liberté esthétique qui aurait été autrement impossible.

**Mots clés**

Créativité assistée par ordinateur, Créativité computationnelle, Composition musicale, Musique à l'image, Sonification d'images, Oulipo.

**Abstract**

Since its conception, the computer has found applications to accompany human creativity. Today, the debate about computers and creativity involves several challenges, such as understanding human creativity, modeling the creative process, and programming the computer to exhibit behavior that appears to be creative to some extent. In this paper, we are interested in how the computer can be used as a tool to promote creativity in a musical composition. We automatically extracted the brightness curve from a silent movie and then used it to compose a piece of music to accompany the movie. We extracted several parameters from the brightness curve, and applied compositional rules from these parameters to write the instrumental music for the film. The final composition has a synchronicity and aesthetic fit with the film that are surprising. This compositional process also allowed for a degree of aesthetic freedom that would otherwise have been impossible.

**Keywords**

Computer-Aided Creativity, Computational Creativity, Musical Composition, Music for Film, Image Sonification, Oulipo.

**Contexte**

La créativité computationnelle (ou CC) est une discipline ayant ses racines dans l'intelligence artificielle, la cybernétique, les sciences cognitives, l'ingénierie, la conception, la psychologie et la philosophie. Elle explore la possibilité que des ordinateurs deviennent des créateurs autonomes à part entière.

En ce qui concerne le domaine de la musique, les tous premiers travaux en informatique musicale / composition assistée par ordinateur remontent aux années 50 et à la fameuse composition par Hiller et Isaacson, la *suite Illiac* (Hiller & Isaacson, 1959).

Il n'existe pas de consensus au sujet des nombreuses questions qui se posent dès lors que l'on discute de créativité assistée par ordinateur (Ritchie, 07), sans compter que la créativité elle-même reste un sujet de débat parmi les chercheurs, les inventeurs et les artistes (Boden, 1996 ; Bentley, 2001 ; Cardoso, 2003 ; Boden, 2004 ; Boden, 2019). La plupart des auteurs (Boden, 2004 ; Duch, 2013) définissent la créativité comme « la capacité de créer des solutions, des artefacts ou des idées qui sont à la fois nouveaux et appropriés, et aussi éventuellement surprenants et utiles »[1]. Cependant, la créativité elle-même échappe encore à une théorie formelle.

Selon Boden (Boden, 2004), la première étape pour répondre à des questions telles que « un ordinateur peut-il être créatif ? » consiste à clarifier ce qu'est la créativité et comment la reproduire. Si certains chercheurs ont consacré des efforts considérables à cette question, d'autres auteurs considèrent qu'elle n'est plus pertinente. Par exemple, Ritchie (Ritchie, 2007) décrit les critères qui peuvent être utilisés pour attribuer la créativité à un programme informatique, tandis que Miranda (Miranda, 2003) déclare que « les ordinateurs peuvent composer s'ils sont programmés de manière appropriée »[2], en parlant spécifiquement de la composition musicale. Néanmoins, de nombreuses autres questions liées à la créativité et aux ordinateurs restent sans réponse. Par exemple, le chapitre 3 de Miranda (2021) discute de la question de savoir non seulement si la musique composée artificiellement peut être considérée comme de la musique réelle, mais aussi si cette musique est découverte ou créée.

Les travaux présentés dans cet article s'inscrivent dans le cadre du projet de recherche IA+CA (Intelligence Artificielle + Création Artistique), qui vise à explorer comment les algorithmes d'intelligence artificielle peuvent aider la création artistique, soit comme une étape complémentaire dans un processus créatif, soit en proposant de nouvelles pistes d'exploration aux artistes. IA+CA propose aussi de porter un regard critique sur le rôle que joue réellement un algorithme d'intelligence artificielle dans un processus créatif. En particulier, nous nous demandons quelle part de cette créativité est due à l'algorithme et quelle part est due au programmeur qui tire des conclusions esthétiques de l'expérience. Plus précisément, nous abordons dans ce projet le domaine de la composition musicale et/ou sonore, ainsi que le domaine des arts visuels, notamment les arts numériques.

Le travail présenté ici est une étape préliminaire, où nous examinons comment certains algorithmes peuvent produire des données comme déclencheurs du geste créatif dans le contexte de la composition musicale pour le cinéma. Plus précisément, la question centrale abordée dans cet article est : « Existe-t-il un moyen de composer de la musique à partir de données extraites d'un film, mais sans nécessairement utiliser le film lui-même comme stimulus à la créativité ? » Nous explorons une façon de le faire qui donne des résultats très intéressants et prometteurs.

---

[1] "the ability to create solutions, artefacts, or ideas that are both novel and appropriate, and also possibly surprising and valuable".
[2] "computers can compose if programmed appropriately."

*Créativité assistée par ordinateur*

Le rôle de l'ordinateur dans le processus créatif figure parmi les questions les plus fondamentales qui sous-tendent les discussions sur la créativité computationnelle[3] (Cardoso, 2003 ; Besold, 2015 ; Veale, 2019). Certains souhaitent explorer le potentiel des ordinateurs à devenir des créateurs autonomes à part entière (Bentley, 2001 ; Veale, 2019), tandis que d'autres considèrent l'ordinateur comme un outil qui aide le créateur ou l'artiste à explorer des possibilités créatives qui seraient hors de sa portée autrement (Todd, 1992). En tant que telle, la créativité computationnelle (Duch, 2013) implique « la capacité de trouver des solutions à la fois nouvelles et appropriées en utilisant des moyens informatiques »[4]. Besold *et al.* (2015) définissent la créativité computationnelle comme « l'utilisation d'ordinateurs pour générer des résultats qui seraient considérés comme créatifs s'ils étaient produits par des humains seuls »[5]. Ici, nous nous intéressons à l'ordinateur comme un outil du point de vue de la composition musicale (Schiavio, 2020, Miranda, 2021).

La créativité exploratoire (Boden, 1996) implique l'exploration et peut-être aussi la transformation des espaces conceptuels dans l'esprit des gens. Les espaces conceptuels sont des styles de pensée structurés. Ce processus de créativité exploratoire nécessite une combinaison équilibrée de contraintes et de liberté créative. Les contraintes sont utilisées pour définir les limites de l'espace conceptuel, tandis que l'artiste conserve suffisamment de liberté créative pour explorer les possibilités à l'intérieur de cet espace conceptuel, avec la capacité de modifier de temps à autre ces contraintes, voire de s'en affranchir. Le calcul évolutif[6] a été largement utilisé pour imiter la créativité exploratoire dans de nombreux domaines artistiques (Bentley, 2001) tels que la musique (Miranda, 2004 ; Miranda, 2007 ; Romero, 2008 ; Briot, 2020) et les arts visuels (Todd, 1992). Inspirée par le mouvement OuLiPo (Oulipo, 1988), notre approche impose des contraintes pour forcer le compositeur à explorer de manière créative l'espace conceptuel résultant.

*Composer de la musique pour le cinéma*

La musique au cinéma peut avoir des fonctions et des modes d'application différents. Cependant, la valeur ajoutée expressive et informative par laquelle les sons enrichissent une image donnée (Chion, 1994) est un aspect fondamental de la musique au cinéma. La composition pour l'image soulève fréquemment des questions liées à la relation entre le film et le son : la composition musicale et sonore doit-elle s'articuler avec les images, ou doit-elle servir de contrepoint aux images et fournir des informations complémentaires ? La recherche de paramètres communs entre l'image et le son peut nous aider à trouver un comportement commun à ces deux langages, à le compléter, ou à créer un contrepoint. Les outils informatiques nous aident à obtenir de nouvelles informations, et nous permettent de trouver des éléments communs entre le langage cinématographique et le langage musical. Par exemple, nous pouvons analyser les images d'un film d'un point de vue musicologique, ou créer une composition musicale avec une approche cinématographique.

**Le processus**

Parmi les nombreuses façons de composer de la musique pour un film, la plus courante consiste à utiliser le film lui-même comme guide une fois qu'il est terminé. Parfois, les compositeurs créent des morceaux de musique avant la fin du film, basés sur le scénario et/ou un dialogue avec le cinéaste. Ici, nous voulons adopter une approche différente : peut-on extraire automatiquement, grâce à l'utilisation d'un algorithme, des paramètres qui peuvent ensuite être utilisés comme guides pour la composition de la musique, en plus ou à la place du film lui-même ?

---

[3] Créativité assistée par ordinateur.
[4] "the capacity to find solutions that are both novel and appropriate using computational means".
[5] "the use of computers to generate results that would be regarded as creative if produced by humans alone".
[6] Evolutionary computation.

*Extraire automatiquement un paramètre des images*

Il existe différentes manières d'extraire automatiquement des paramètres d'images ou de films. Le but est généralement de réduire la dimensionnalité des données, de faciliter leur manipulation et/ou leur interprétation. En effet, une séquence d'images 2D peut contenir une quantité écrasante d'informations. Pour réduire cette dimensionnalité des données, nous pouvons extraire un seul paramètre qui peut ensuite être la base du geste créatif. Dans le cadre de la créativité, le but est presque toujours de créer les conditions de la nouveauté, d'apporter de nouvelles contraintes, afin de favoriser une nouvelle approche, une nouvelle façon de faire, déclenchant ainsi la créativité. En ce sens, nous nous inscrivons complètement dans la tradition oulipienne, c'est-à-dire dans la recherche de contraintes ou de règles qui servent de guides, et même de stimulus, au geste créatif (Oulipo, 1988).

*Choix des paramètres*

Les paramètres de composition doivent être interprétables et extractibles. « Interprétables » signifie qu'ils doivent transmettre au compositeur des informations utilisables et pertinentes par rapport à la tâche à accomplir, ici composer une musique de film. « Extractibles » signifie simplement que nous devons pouvoir extraire automatiquement ce paramètre avec un algorithme. Pour ce travail, nous avons décidé d'utiliser la courbe de luminosité comme paramètre de composition, car elle est à la fois extractible et interprétable. La raison derrière cela était que l'un de nous avait déjà travaillé avec ce paramètre et avait déjà quelques connaissances et quelques programmes en Max MSP pour l'extraire.

*Travail précédent*

La composition *Outer Space*, commandée à Javier Elipe par l'Ircam - Centre Pompidou en 2019, a utilisé différents algorithmes d'analyse d'images du film *Outer Space* de Peter Tscherkassky (1999) pour extraire des paramètres communs entre le film et la composition musicale. L'utilisation de ces algorithmes avait pour but de créer une articulation entre la musique et les images, qui soit adaptée à chaque moment du film, et en même temps évolue au fur et à mesure du film.

La bibliothèque Jitter du logiciel Max/MSP a été utilisée pour analyser l'évolution de la luminosité d'un film à l'échelle de l'image tout entière, mais aussi au niveau de certaines zones de l'image. La courbe de luminosité, extraite à l'échelle de l'image tout entière, a été utilisée à différentes fins, telles que l'interprétation de la structure générale du film, la distribution de l'énergie entre les plans visuel et sonore, et l'utilisation de cette courbe comme cinquième instrument.

*Composer en utilisant uniquement la courbe de luminosité*

La courbe de luminosité constitue une donnée intéressante pour la composition. Il existe de nombreuses interprétations liées au fait qu'une image est claire ou sombre. Nous utilisons même les termes « clair » et « sombre » comme métaphores dans notre langage quotidien. Nous avons également beaucoup d'associations esthétiques avec la luminosité et l'obscurité. Et même en esthétique visuelle, nous avons tendance à manipuler la luminosité et l'obscurité en fonction du type d'émotions que nous voulons susciter chez le spectateur.

Pour cette expérience préliminaire, nous avons décidé d'utiliser uniquement la courbe de luminosité, sans nous appuyer du tout sur le film lui-même, car la courbe de luminosité nous a semblé être une représentation appropriée du film qui transmet suffisamment d'informations interprétatives pour composer de la musique pour celui-ci.

*Une nouvelle façon d'aborder la composition*

Un aspect particulièrement intéressant de cette approche est qu'elle oblige le compositeur à explorer de nouvelles esthétiques en adoptant un processus de composition qui lui est nouveau et étranger. L'utilisation de la courbe de luminosité pour composer de la musique est une façon de définir une certaine contrainte qui délimite un espace conceptuel à l'intérieur duquel la composition peut prendre place, un peu dans la lignée du mouvement OuLiPo (Oulipo, 1988) qui proposait de créer des règles et des contraintes arbitraires afin de déclencher la créativité.

**La composition**

Une fois que nous avons choisi la courbe de luminosité comme paramètre à utiliser, nous avons également dû choisir le film à partir duquel l'extraire. Le choix du paramètre a un impact sur le choix du film (et vice versa). Le choix de la courbe de luminosité nécessite un film présentant une variation dynamique du degré de luminosité suffisante pour permettre au compositeur d'en extraire des informations esthétiques. Par exemple, la courbe de luminosité d'un épisode d'*Utopia* (2013) coïncide principalement avec les emplacements des coupures entre les plans, ce qui ne serait pas très pertinent pour composer de la musique (voir figure 1).

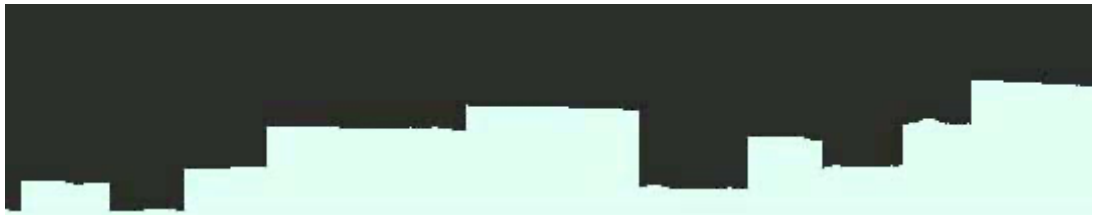
Figure 1 : Courbe de luminosité pour une séquence du film *Utopia* (2013).

Elle est beaucoup plus pertinente lorsqu'on l'applique à des films expérimentaux. Nous avons finalement choisi de travailler sur 1 minute et 30 secondes du film *Rhythmus 21* de Hans Richter (1921) (voir figure 2).

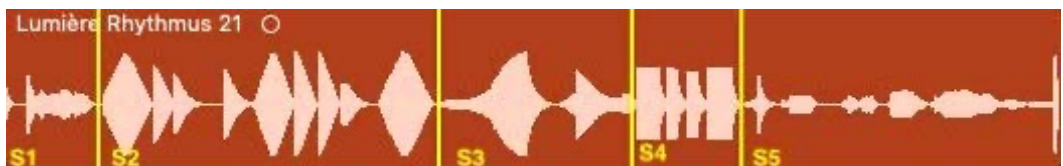
Figure 2 : Division de la courbe de luminosité de *Rhythmus 21* de Hans Richter (1921)
en sections pour faciliter la composition.

*Sections et choix de l'échelle*

Afin de guider le processus de composition, nous avons également choisi de diviser le segment choisi en sections, car elles avaient un sens sur la base de la courbe de luminosité (voir figure 2). Cette segmentation est bien sûr subjective, mais elle a du sens sur la base de la courbe. L'échelle est également un facteur important ici. Si nous faisons un zoom avant, nous aurons plus de détails et plus d'incitations à la créativité, mais les changements seront trop rapides à l'échelle temporelle. Si nous effectuons un zoom arrière, nous aurons davantage d'unité dans la composition, mais moins d'incitations à la créativité, et la composition ne sera pas aussi intéressante. Les sections indiquées à la figure 2 constituent un bon compromis.

*Interprétation de la courbe de luminosité*

L'interprétation de la courbe nous laisse initialement une certaine liberté, mais une analyse plus détaillée de son évolution et de sa structure globale peut nous fournir d'autres informations, qui ne sont pas appréciables au départ. Les différentes évolutions de la luminosité suggérées par la courbe peuvent présenter différents degrés de symétrie : par exemple, des figures géométriques identifiables (triangles, carrés, etc.), ou des formes qui ne représentent aucune figure géométrique. Nous pouvons également apprécier différents types de linéarité : par exemple, des figures géométriques avec des lignes droites, ou avec des lignes ondulées (voir figure 3).

Une analyse de ces évolutions de la luminosité à travers le film sera importante pour une interprétation musicale ou sonore de la courbe de luminosité. Nous pourrions même envisager que ces évolutions de la courbe de luminosité soient en quelque sorte des « gestes de luminosité », que nous essayerions alors de faire coïncider avec des gestes de composition. Par exemple, la même figure géométrique (ou très similaire) dans différentes parties du film peut être interprétée comme un geste récurrent, même si les images qui correspondent à ces gestes ne sont pas totalement similaires. Cela peut signifier que l'interprétation de ces « gestes de luminosité » peut nous donner un deuxième niveau d'interprétation de la forme du film et de la composition sonore/musicale.

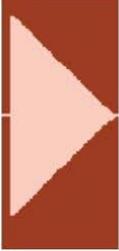
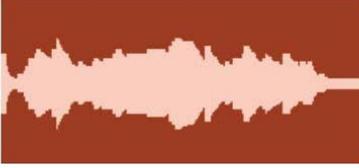
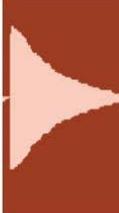
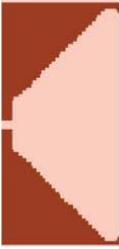
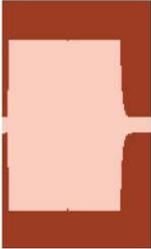
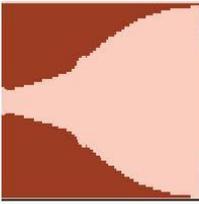

Figure 3 : Exemples de règles de composition
à partir de la courbe de luminosité de *Rhythmus 21* de Hans Richter (1921).

Un autre paramètre d'interprétation peut être l'intelligibilité (ou la reconnaissance) de formes trouvées dans l'analyse des courbes. Des figures reconnaissables pourraient conduire à l'interprétation de fragments similaires comme un geste musical caractéristique. D'un autre côté, une courbe ondulante pourrait être interprétée comme une texture musicale complexe sans gestes musicaux caractéristiques. Enfin, l'amplitude de la courbe de luminosité peut être mise en correspondance avec des caractéristiques timbrales, où une luminosité élevée évoque des timbres brillants et des passages sombres évoquent des passages plus ternes et plus silencieux.

*Suivre les règles ou non*

L'interprétation de la courbe de luminosité impose une contrainte tout en laissant la liberté d'utiliser ces règles auto-imposées en fonction des besoins de la composition. De même, les règles elles-mêmes peuvent évoluer au fur et à mesure que l'on réfléchit à la composition musicale/sonore. Enfin, il est important de trouver un ensemble de règles et d'interprétations de la courbe qui assurent la cohérence entre la composition musicale en cours et une interprétation formelle et gestuelle de la courbe d'analyse de la luminosité. Cela nous permettra finalement de trouver une série de concepts d'interprétation musicale/formelle communs au niveau cinématographique et musical.

*Résultats*

Le résultat de cette expérience peut être vu et entendu en suivant ce lien[7]. La première vidéo montre les premières 1m30s du film avec la musique composée uniquement à partir de la courbe de luminosité. La deuxième vidéo montre la musique composée uniquement à partir de la courbe de luminosité ainsi que cette courbe qui a servi d'incitation à la création de la musique. La troisième vidéo montre la musique avec le film et la courbe de luminosité.

---

[7] https://www.ivanhereandnow.com/rhythmus21

La courbe de luminosité permet plusieurs interprétations et utilisations. Pour cette expérience, nous avons utilisé la courbe pour analyser le film de manière globale : les différentes formes géométriques et textures visuelles que nous avons pu trouver dans la courbe de luminosité nous ont permis d'apprécier différentes sections et structures. Ce type d'analyse pourrait sembler un peu aléatoire, mais après les différentes expériences, cette analyse formelle des structures visuelles correspond bien à une distribution des énergies visuelles et sonores.

Mais la courbe de luminosité nous a aussi renseignés sur la relation entre image et musique : la courbe de luminosité a pu aussi être utilisée pour faire une « traduction » musicale de cette courbe. Ce procédé peut sembler redondant et évident. Mais les expériences que nous avons menées nous ont montré que cette relation n'était pas si évidente. Par exemple, un triangle dans la courbe de luminosité (équivalent par exemple à un crescendo musical) peut présenter une montée progressive dans le film qui n'est pas vraiment perçue comme un crescendo. Le résultat final avec la musique a une certaine logique de comportement, et n'est pas perçu comme une imitation de la courbe.

Figure 4 : Quelques extraits de la partition
réalisée par Felipe Ariani pour enrichir la composition
basée sur la courbe de luminosité de *Rhythmus 21* de Hans Richter (1921).

**La partition**

La composition musicale réalisée avec la courbe de luminosité a été composée avec le logiciel *Logic* et l'utilisation d'instruments virtuels. A priori, ce travail ne nécessite pas de partition musicale, mais la création d'une partition et le travail d'édition musicale peuvent apporter des éléments supplémentaires à l'interprétation de la courbe (voir figure 4).

Même si nous pouvons utiliser une grande variété d'articulations et de samplers avec les instruments virtuels, la partition peut nous donner quelques éléments supplémentaires sur les gestes instrumentaux. Par exemple, écrire une ligne graphique détaillée près de la courbe de luminosité qui correspond au passage de *sul ponticello* à *sul tanto* nous permet d'utiliser une gestualité de mouvement d'archet similaire à l'analyse de la luminosité. De même, dans une partition, on peut faire des évolutions très précises de la vitesse du trémolo afin d'approcher les différentes granulations de la courbe de luminosité. Ces types de détails sont difficiles à traiter avec des instruments virtuels.

D'autre part, la partition peut nous permettre d'explorer avec un interprète des modes de jeu qui sont difficiles à trouver sur un instrument virtuel. Enfin, la dimension visuelle d'une partition et l'utilisation d'éléments graphiques dans l'écriture instrumentale nous apportent un certain degré d'organicité, qui peut être perdu dans une composition musicale réalisée uniquement avec des outils informatiques.

**Travaux futurs**

Ces résultats très prometteurs nous encouragent à aller plus loin.

*Autres paramètres*

La première piste à explorer est d'utiliser d'autres paramètres que la courbe de luminosité. Il existe de nombreux paramètres qui peuvent être extraits automatiquement d'images en mouvement. Une extension très simple de la courbe de luminosité serait d'extraire une courbe de luminosité pour chaque couleur de capteur, rouge, vert et bleu, et de voir ensuite si les courbes de luminosité des couleurs déclenchent des gestes créatifs différents. Le contraste pourrait également être utilisé à la place de la luminosité, et permettrait de voir comment les pixels les plus clairs d'une image se rapportent aux pixels les plus sombres. Finalement, l'objectif est de passer progressivement à des algorithmes d'intelligence artificielle pour aider à déterminer quels types de paramètres peuvent être extraits automatiquement et dynamiquement d'images en mouvement, en termes de pertinence, pour être utilisés comme déclencheurs d'une composition musicale.

*Autres films*

Un autre prolongement de cette étude serait de tester cette approche, ainsi que d'autres approches, sur d'autres films. Nous avons établi précédemment que certains films (les plus expérimentaux) étaient plus appropriés pour ce type d'expérimentation. Mais nous aurions besoin d'explorer davantage afin de décider quelles caractéristiques d'un film sont pertinentes pour quels paramètres extraits.

*Autres règles*

La courbe de luminosité pourrait également nous permettre de travailler sur d'autres paramètres. Par exemple :
- un principe de complémentarité avec la partie visuelle, que nous appelons « orchestration audiovisuelle » : il s'agirait d'utiliser la courbe pour répartir l'énergie audiovisuelle du projet, afin d'éviter l'utilisation systématique d'une texture musicale très lourde avec une courbe de luminosité très vive, ou l'inverse.
- donner un contrepoint rythmique à l'énergie donnée par l'image : avec l'utilisation d'un granulateur dans Max/MSP, nous pourrions obtenir un rythme parallèle à celui utilisé par les images, obtenant ainsi un contrepoint entre images et musique, tout en utilisant une certaine cohérence produite par l'analyse de la luminosité.

*Utiliser la partition*

L'utilisation de la partition et pas seulement d'une simulation MIDI nous permettrait également de travailler sur les gestes instrumentaux avec les différents gestes de luminosité. La réalisation de la partition au cours de cette expérience nous a permis d'obtenir des éléments d'une plus grande précision, que nous ne pouvions pas obtenir avec une maquette MIDI.

*Automatiser ce qui peut être automatisé*

L'intérêt de ce type d'approche est de pouvoir automatiser ce qui peut l'être. On recherche donc, dans les étapes d'un processus créatif, ici de composition musicale, des étapes qui pourraient être confiées à un algorithme (par exemple un algorithme d'intelligence artificielle), de manière à conduire une exploration plus systématique des différentes possibilités, ce qui prendrait un temps beaucoup plus long sans algorithme. L'idée de cette approche par automatisation est toujours de libérer le compositeur au maximum pour qu'il puisse se concentrer sur ses gestes créatifs, tout en utilisant l'automatisation pour délimiter un nouvel espace de création.

**Conclusion**

Il existe de nombreuses questions sans réponse concernant l'application de l'IA dans les arts (Machado, 2021), et plus particulièrement dans la musique (Miranda, 2021). Aujourd'hui, la plupart des efforts de recherche sont consacrés au développement d'algorithmes capables de créer de la musique (Miranda, 2021). Par exemple, dans le chapitre 4 de Machado 2021, Wiggins discute des questions philosophiques liées à l'intelligence musicale artificielle, tandis que le chapitre de Gioti dans Miranda 2021 explore le potentiel de l'IA pour façonner les idées de composition. Ce débat tourne autour du rôle de l'ordinateur, et plus particulièrement de l'IA, dans le processus de création. L'ordinateur est-il censé aider ou remplacer l'artiste ? L'automatisation est-elle vraiment bénéfique à la créativité ? Si oui, comment l'automatisation peut-elle favoriser la créativité plutôt que de l'entraver ? Le projet IA+CA vise à étudier certaines de ces questions ouvertes à l'intersection de la musique et des arts visuels. Bien que préliminaires, les travaux décrits ici ont abordé nombre de ces questions, ce qui nous encourage à continuer d'explorer le rôle de l'IA dans la conception sonore, la composition musicale et les arts visuels.

*Contributions des auteurs*

Tous les auteurs ont participé à la rédaction de ce manuscrit et aux expériences, et sont énumérés par ordre alphabétique.